\documentclass[%
 reprint,
 twocolumn,
superscriptaddress,
nofootinbib,
 amsmath,amssymb,
 prd,
]{revtex4-2}

\usepackage{graphicx}% Include figure files
\usepackage{dcolumn}% Align table columns on decimal point
\usepackage{bm}% bold math
\usepackage{color}
\usepackage{siunitx}
\usepackage[colorlinks=true,citecolor=blue,urlcolor=blue]{hyperref}
\usepackage[dvipsnames]{xcolor}
\usepackage{lineno}
\usepackage{bbm}
\usepackage{orcidlink}
\usepackage{url}
\usepackage{ulem} %Remove me after editing

\newcommand*{\diff}{\,\mathrm{d}}
\newcommand*{\eexp}{\mathrm{e}}

\renewcommand{\eqref}[1]{Eq.~(\ref{#1})}
\newcommand{\eqsref}[2]{Eqs.~(\ref{#1}) and (\ref{#2})}
\newcommand{\figref}[1]{Fig.~\ref{#1}}

\newcommand{\secref}[1]{Sec.~\ref{#1}}
\newcommand{\appref}[1]{Appendix~\ref{#1}}

\newcommand{\Romann}[1]{\MakeUppercase{\romannumeral #1}}
\newcommand{\gstlal}{GstLAL}
\usepackage{acronym}
\begin{document}

\preprint{APS/123-QED}

\title{Improved ranking statistics of the \gstlal{} inspiral search\\for compact binary coalescences}% Force line breaks with \\

\author{Leo Tsukada \orcidlink{0000-0003-0596-5648}}
\email{leo.tsukada@ligo.org}
\affiliation{Department of Physics, The Pennsylvania State University, University Park, PA 16802, USA}
\affiliation{Institute for Gravitation and the Cosmos, The Pennsylvania State University, University Park, PA 16802, USA}

\author{Prathamesh Joshi \orcidlink{0000-0002-4148-4932}}
\email{prathamesh.joshi@ligo.org}
\affiliation{Department of Physics, The Pennsylvania State University, University Park, PA 16802, USA}
\affiliation{Institute for Gravitation and the Cosmos, The Pennsylvania State University, University Park, PA 16802, USA}

\author{Shomik Adhicary}
\affiliation{Department of Physics, The Pennsylvania State University, University Park, PA 16802, USA}
\affiliation{Institute for Gravitation and the Cosmos, The Pennsylvania State University, University Park, PA 16802, USA}

\author{Richard George  \orcidlink{0000-0002-7797-7683}}
\affiliation{Center for Gravitational Physics, University of Texas at Austin, Austin, TX 78712, USA}

\author{Andre Guimaraes \orcidlink{0000-0002-0648-4494}}
\affiliation{Department of Physics \& Astronomy, Louisiana State University, Baton Rouge, LA 70803, USA}

\author{Chad Hanna}
\affiliation{Department of Physics, The Pennsylvania State University, University Park, PA 16802, USA}
\affiliation{Institute for Gravitation and the Cosmos, The Pennsylvania State University, University Park, PA 16802, USA}
\affiliation{Department of Astronomy and Astrophysics, The Pennsylvania State University, University Park, PA 16802, USA}
\affiliation{Institute for Computational and Data Sciences, The Pennsylvania State University, University Park, PA 16802, USA}

\author{Ryan Magee \orcidlink{0000-0001-9769-531X}}
\affiliation{LIGO Laboratory, California Institute of Technology, Pasadena, CA 91125, USA}

\author{Aaron Zimmerman}
\affiliation{Center for Gravitational Physics, University of Texas at Austin, Austin, TX 78712, USA}

\author{Pratyusava Baral \orcidlink{0000-0001-6308-211X}}
\affiliation{Leonard E.\ Parker Center for Gravitation, Cosmology, and Astrophysics, University of Wisconsin-Milwaukee, Milwaukee, WI 53201, USA}

\author{Amanda Baylor \orcidlink{0000-0003-0918-0864}}
\affiliation{Leonard E.\ Parker Center for Gravitation, Cosmology, and Astrophysics, University of Wisconsin-Milwaukee, Milwaukee, WI 53201, USA}

\author{Kipp Cannon \orcidlink{0000-0003-4068-6572}}
\affiliation{RESCEU, The University of Tokyo, Tokyo, 113-0033, Japan}

\author{Sarah Caudill}
\affiliation{Nikhef, Science Park, 1098 XG Amsterdam, Netherlands}

\author{Bryce Cousins \orcidlink{0000-0002-7026-1340}}
\affiliation{Department of Physics, University of Illinois, Urbana, IL 61801 USA}
\affiliation{Department of Physics, The Pennsylvania State University, University Park, PA 16802, USA}
\affiliation{Institute for Gravitation and the Cosmos, The Pennsylvania State University, University Park, PA 16802, USA}

\author{Jolien D. E. Creighton \orcidlink{0000-0003-3600-2406}}
\affiliation{Leonard E.\ Parker Center for Gravitation, Cosmology, and Astrophysics, University of Wisconsin-Milwaukee, Milwaukee, WI 53201, USA}

\author{Becca Ewing}
\affiliation{Department of Physics, The Pennsylvania State University, University Park, PA 16802, USA}
\affiliation{Institute for Gravitation and the Cosmos, The Pennsylvania State University, University Park, PA 16802, USA}

\author{Heather Fong}
\affiliation{RESCEU, The University of Tokyo, Tokyo, 113-0033, Japan}
\affiliation{Graduate School of Science, The University of Tokyo, Tokyo 113-0033, Japan}

\author{Patrick Godwin}
\affiliation{LIGO Laboratory, California Institute of Technology, Pasadena, CA 91125, USA}
\affiliation{Department of Physics, The Pennsylvania State University, University Park, PA 16802, USA}
\affiliation{Institute for Gravitation and the Cosmos, The Pennsylvania State University, University Park, PA 16802, USA}

\author{Reiko Harada}
\affiliation{RESCEU, The University of Tokyo, Tokyo, 113-0033, Japan}
\affiliation{Graduate School of Science, The University of Tokyo, Tokyo 113-0033, Japan}

\author{Yun-Jing Huang \orcidlink{0000-0002-2952-8429}}
\affiliation{Department of Physics, The Pennsylvania State University, University Park, PA 16802, USA}
\affiliation{Institute for Gravitation and the Cosmos, The Pennsylvania State University, University Park, PA 16802, USA}

\author{Rachael Huxford}
\affiliation{Department of Physics, The Pennsylvania State University, University Park, PA 16802, USA}
\affiliation{Institute for Gravitation and the Cosmos, The Pennsylvania State University, University Park, PA 16802, USA}

\author{James Kennington \orcidlink{0000-0002-6899-3833}}
\affiliation{Department of Physics, The Pennsylvania State University, University Park, PA 16802, USA}
\affiliation{Institute for Gravitation and the Cosmos, The Pennsylvania State University, University Park, PA 16802, USA}

\author{Soichiro Kuwahara}
\affiliation{RESCEU, The University of Tokyo, Tokyo, 113-0033, Japan}
\affiliation{Graduate School of Science, The University of Tokyo, Tokyo 113-0033, Japan}

\author{Alvin K. Y. Li \orcidlink{0000-0001-6728-6523}}
\affiliation{LIGO Laboratory, California Institute of Technology, Pasadena, CA 91125, USA}

\author{Duncan Meacher \orcidlink{0000-0001-5882-0368}}
\affiliation{Leonard E.\ Parker Center for Gravitation, Cosmology, and Astrophysics, University of Wisconsin-Milwaukee, Milwaukee, WI 53201, USA}

\author{Cody Messick}
\affiliation{MIT Kavli Institute for Astrophysics and Space Research, Massachusetts Institute of Technology, Cambridge, MA 02139, USA}

\author{Soichiro Morisaki \orcidlink{0000-0002-8445-6747}}
\affiliation{Institute for Cosmic Ray Research, The University of Tokyo, 5-1-5 Kashiwanoha, Kashiwa, Chiba 277-8582, Japan}
\affiliation{Leonard E.\ Parker Center for Gravitation, Cosmology, and Astrophysics, University of Wisconsin-Milwaukee, Milwaukee, WI 53201, USA}

\author{Debnandini Mukherjee  \orcidlink{0000-0001-7335-9418}}
\affiliation{NASA Marshall Space Flight Center, Huntsville, AL 35811, USA}
\affiliation{Center for Space Plasma and Aeronomic Research, University of Alabama in Huntsville, Huntsville, AL 35899, USA}

\author{Wanting Niu}
\affiliation{Department of Physics, The Pennsylvania State University, University Park, PA 16802, USA}
\affiliation{Institute for Gravitation and the Cosmos, The Pennsylvania State University, University Park, PA 16802, USA}

\author{Alex Pace}
\affiliation{Department of Physics, The Pennsylvania State University, University Park, PA 16802, USA}
\affiliation{Institute for Gravitation and the Cosmos, The Pennsylvania State University, University Park, PA 16802, USA}

\author{Cort Posnansky}
\affiliation{Department of Physics, The Pennsylvania State University, University Park, PA 16802, USA}
\affiliation{Institute for Gravitation and the Cosmos, The Pennsylvania State University, University Park, PA 16802, USA}

\author{Anarya Ray}
\affiliation{Leonard E.\ Parker Center for Gravitation, Cosmology, and Astrophysics, University of Wisconsin-Milwaukee, Milwaukee, WI 53201, USA}

\author{Surabhi Sachdev \orcidlink{0000-0002-0525-2317}}
\affiliation{School of Physics, Georgia Institute of Technology, Atlanta, GW 30332, USA}
\affiliation{Leonard E.\ Parker Center for Gravitation, Cosmology, and Astrophysics, University of Wisconsin-Milwaukee, Milwaukee, WI 53201, USA}

\author{Shio Sakon \orcidlink{0000-0002-5861-3024}}
\affiliation{Department of Physics, The Pennsylvania State University, University Park, PA 16802, USA}
\affiliation{Institute for Gravitation and the Cosmos, The Pennsylvania State University, University Park, PA 16802, USA}

\author{Divya Singh \orcidlink{0000-0001-9675-4584}}
\affiliation{Department of Physics, The Pennsylvania State University, University Park, PA 16802, USA}
\affiliation{Institute for Gravitation and the Cosmos, The Pennsylvania State University, University Park, PA 16802, USA}

\author{Ron Tapia}
\affiliation{Department of Physics, The Pennsylvania State University, University Park, PA 16802, USA}
\affiliation{Institute for Computational and Data Sciences, The Pennsylvania State University, University Park, PA 16802, USA}

\author{Takuya Tsutsui \orcidlink{0000-0002-2909-0471}}
\affiliation{RESCEU, The University of Tokyo, Tokyo, 113-0033, Japan}

\author{Koh Ueno \orcidlink{0000-0003-3227-6055}}
\affiliation{RESCEU, The University of Tokyo, Tokyo, 113-0033, Japan}

\author{Aaron Viets \orcidlink{0000-0002-4241-1428}}
\affiliation{Concordia University Wisconsin, Mequon, WI 53097, USA}

\author{Leslie Wade}
\affiliation{Department of Physics, Hayes Hall, Kenyon College, Gambier, Ohio 43022, USA}

\author{Madeline Wade \orcidlink{0000-0002-5703-4469}}
\affiliation{Department of Physics, Hayes Hall, Kenyon College, Gambier, Ohio 43022, USA}

\date{\today}% It is always \today, today,
             %  but any date may be explicitly specified

\begin{abstract}
    Starting from May 2023, the LIGO Scientific, Virgo and KAGRA Collaboration
    is planning to conduct the fourth observing run with improved detector
    sensitivities and an expanded detector network including KAGRA. Accordingly,
    it is vital to optimize the detection algorithm of low-latency search
    pipelines, increasing their sensitivities to gravitational waves from
    compact binary coalescences. In this work, we discuss several new features
    developed for ranking statistics of \gstlal{}-based inspiral pipeline, which
    mainly consist of: the signal contamination removal, the bank-$\xi^2$
    incorporation, the upgraded $\rho-\xi^2$ signal model and the integration of
    KAGRA.  An injection study demonstrates that these new features improve the
    pipeline's sensitivity by approximately \SIrange{15}{20}{\percent}, paving
    the way to further multi-messenger observations during the upcoming
    observing run.
\end{abstract}

%\keywords{Suggested keywords}%Use showkeys class option if keyword
                              %display desired
\maketitle

%% Acronyms
\acrodef{gw}[GW]{gravitational wave}
\acrodef{bh}[BH]{black hole}
\acrodef{cbc}[CBC]{compact binary coalescence}
\acrodef{bbh}[BBH]{binary black hole}
\acrodef{bns}[BNS]{binary neutron star}
\acrodef{imbh}[IMBH]{intermediate-mass black hole}
\acrodef{ligo}[LIGO]{the Laser Interferometer Gravitational-wave Observatory}
\acrodef{lvk}[LVK]{the LIGO Scientific, Virgo and KAGRA Collaboration}
% \acrodef{lr}[LR]{likelihood ratio}
\acrodef{o2}[O2]{the second observing run}
\acrodef{o3}[O3]{the third observing run}
\acrodef{o3a}[O3a]{the first half of the third observing run}
\acrodef{o4}[O4]{the fourth observing run}
\acrodef{o5}[O5]{the fifth observing run}
\acrodef{snr}[SNR]{signal-to-noise ratio}
\acrodef{csd}[CSD]{cross spectral density}
\acrodef{psd}[PSD]{power spectral density}
\acrodef{pdf}[PDF]{probability density function}
\acrodef{gwtc}[GWTC]{Gravitational Wave Transient Catalog}
\acrodef{vt}[$VT$]{sensitive space-time volume}
\acrodef{far}[FAR]{false alarm rate}
%\tableofcontents

\section{Introduction}
The recent consistent detections~\cite{gwtc-1, gwtc-2, gwtc-2.1, gwtc-3} of
\acp{gw} by \ac{ligo}~\cite{ligo} and Virgo~\cite{virgo} have opened up a new
window to observe the Universe and established the field of \ac{gw} astronomy,
allowing us to study some of the most energetic events in the Universe, e.g. the
mergers of \acp{bbh} and \acp{bns}. In particular, the discovery of a \ac{gw} signal from
a \ac{bns}, GW170817~\cite{170817}, enabled the subsequent detection of
electromagnetic counterparts across a broad frequency spectrum, including
gamma-rays, X-rays, optical light, and radio waves.~\cite{170817_mm} This
real-time observation of GW170817 was made possible by the low-latency detection
pipelines to search for \ac{gw}s. The \gstlal-based inspiral pipeline (referred
to as \gstlal{} hereafter), in particular, played a significant role in
detecting GW170817, as it identified the signal in low
latency~\cite{170817_gcn}, which allowed for a public alert sent out to external
facilities.  The multi-messenger observation, involving both gravitational and
electromagnetic radiation has enriched our understanding of nuclear physics and
astrophysical processes involved in such mergers~\cite{170817_eos,
170817_ejecta, 170817_eos, 170817_pe, 170817_progen}.

Starting from May 2023, \ac{lvk} is planning to conduct
\ac{o4}~\cite{o4date} with the improved detector sensitivities and an expanded
detector network including KAGRA~\cite{kagra}.  Therefore, it is increasingly
important to continue improving the detection efficiency of the pipeline, so
that we can identify more of GW170817-like events in real time and make more
multi-messenger discoveries.

\gstlal{}~\cite{gstlal_cody, gstlal_o2, dtdphi} is a \ac{gw} detection pipeline
that processes the strain data from ground-based \ac{gw} detectors to search for
\ac{gw} signals in low latency. The \gstlal{} library~\cite{gstlal_lib} consists
of a collection of GStreamer libraries~\cite{gstreamer} and plug-ins that depend
on the \ac{ligo} Algorithm Library, LALSuite~\cite{lalsuite}. One of the key
techniques of \gstlal{} is the matched-filtering algorithm~\cite{Wainstein1970,
kip_mf, Sathyaprakash:1991aa, PhysRevLett.70.2984, Finn:1992aa, Finn:1993aa,
PhysRevD.49.1707, Balasubramanian:1996aa, PhysRevD.57.4535}, which involves
comparing observed strain data to a group of theoretical waveforms (referred to
as \textit{a template bank}~\cite{deb_template_bank, shio_template_bank}) that
describes the expected \ac{gw} signals from \acp{cbc}.  This technique has been
widely used in several detection pipelines such as PyCBC~\cite{pycbc},
MBTA~\cite{mbta} and spiir~\cite{spiir}.

After the pipeline identifies \ac{gw} candidates, it assigns significance to
each of them to make a statistical statement about the detections.  The
definition of this significance assignment is different across detection
pipelines and one of the unique features of \gstlal{} is its use of a likelihood
ratio as a ranking statistic~\cite{kipp_lr}. This statistic measures
the probability of the data being produced by a \ac{gw} signal
compared to noise alone, and is a powerful tool for identifying weak signals
buried in noise.

This work presents the recent developments in the \gstlal's detection algorithm,
with a particular focus on the likelihood ratio statistic. First, in
\secref{sec:lr} we review an overview of \gstlal's ranking statistics.  Second,
in \secref{sec:dev} we describe several new features developed in the likelihood
ratio calculation, which are implemented in the analysis configuration for
\ac{o4}.  \secref{sec:results} illustrates the improvement of the detection
sensitivity due to these new features, using a simulated injection analysis.
Lastly, in \secref{sec:concl} we summarize our findings and future work
regarding the developments described here.

\section{Ranking Statistic}
\label{sec:lr}

Since the era of the first observing run, \gstlal{} has adopted the likelihood
ratio as the ranking statistic to evaluate the significance of \ac{gw}
candidates \cite{kipp_lr}.  According to the Neyman-Pearson lemma
\cite{neyman_pearson}, the use of the likelihood ratio is known to provide the
most powerful statistical test at a fixed false alarm probability. Here we
summarize the likelihood ratio implemented in \gstlal{} with particular notes
for the terms relevant to our development toward \ac{o4}.

In \gstlal, the likelihood ratio takes the form of \cite{kipp_lr}
\begin{align}
    \label{eq:lr}
    \mathcal{L}=\frac{P\left(\vec{O}, \vec{\rho}, \vec{\xi^2}, \vec{t}, \vec{\phi}, \theta \mid \mathcal{H}_\mathrm{s}\right)}{P\left(\vec{O}, \vec{\rho}, \vec{\xi^2}, \vec{t}, \vec{\phi}, \theta \mid \mathcal{H}_\mathrm{n}\right)},
\end{align}
which represents the probability of obtaining a set of observable parameters
under the signal hypothesis ($\mathcal{H}_\mathrm{s}$) relative to that under
the noise hypothesis ($\mathcal{H}_\mathrm{n}$). Each vector quantity in
\eqref{eq:lr} denotes observable parameters specific to \ac{gw} detectors.
$\vec{O}$ is a subset of the $N$ detectors detecting an event in coincidence.
$\vec{\rho}$ and $\vec{\xi^2}$ are the vectorized \acp{snr} and
$\xi^2$-signal-based-veto parameter (see \secref{sec:combochisq}) respectively,
measured by each of these detectors.  Similarly, $\vec{t}$ ($\vec{\phi}$) are
the vectorized event's time (phases) at the coalescence on each detector's
frame.  Note that $\theta$ represents four template parameters $\theta=\{m_1,
m_2, s_{1,z}, s_{2,z}\}$, where $m_i,s_{i,z}$ are $i$-th component mass such
that $m_1\geq m_2$, and $i$-th spin component along with the orbital angular momentum
vector of a binary system, respectively. These parameters are unique across the
entire template bank, and hence $\theta$ also serves as a template identifier.
\subsection{Signal model}
The \ac{pdf} in the numerator of \eqref{eq:lr} describes our assumption or prior
knowledge about the observable parameters in the presence of \ac{gw} signals.
We factorize this whole \ac{pdf} as follows \cite{dtdphi}
\begin{align}
    \label{eq:signal_model}
    \begin{aligned}
        P(\ldots \mid \mathcal{H}_\mathrm{s}) & =P\left(\theta \mid \mathcal{H}_\mathrm{s}\right) \\
        & \times P\left(t_\mathrm{ref}, \phi_\mathrm{ref} \mid \theta, \mathcal{H}_\mathrm{s}\right) \\
        & \times P\left(\vec{O} \mid t_\mathrm{ref}, \mathcal{H}_\mathrm{s}\right) \\
        & \times P\left(\vec{\rho}, \vec{\Delta t}, \vec{\Delta\phi} \mid \vec{O}, t_\mathrm{ref}, \mathcal{H}_\mathrm{s}\right)\\
        & \times P\left(\vec{\xi^2} \mid \vec{\rho}, \theta, \mathcal{H}_\mathrm{s}\right),
    \end{aligned}
\end{align}
where $\vec{\Delta t} = \vec{t} - t_\mathrm{ref}$, $\vec{\Delta\phi} =
\vec{\phi} - \phi_\mathrm{ref}$ are the time and phase vectors with reduced
dimensions, being defined relative to a reference detector. Some of the
conditioned parameters are assumed to be independent of others and so do not
appear in every component of the factorized \ac{pdf}. We discuss each component
of the factorized \ac{pdf} in what follows.

The term $P\left(\theta \mid \mathcal{H}_\mathrm{s}\right)$ denotes our model of
how likely each template is to recover a \ac{gw} signal, the so-called \textit{population
model}. This model encodes our prior knowledge about source distributions in a
template bank, as well as the effect of noise fluctuation pushing the point
estimate from the true template to another~\cite{heather_thesis}.  Refer to
\cite{shio_template_bank} for more details of the population model we use for
\ac{o4}.

For a given template, we track its horizon distance for
each operating detector, which can vary over time due to subtle change in the
detector configuration during its observation.  
This allows us to compute
$P(t_\mathrm{ref}, \phi_\mathrm{ref} \mid \theta,
\mathcal{H}_\mathrm{s})$, which scales with the observable volume for
isotropically distributed \ac{gw} sources
\begin{align}
    P\left(t_\mathrm{ref}, \phi_\mathrm{ref} \mid \theta, \mathcal{H}_\mathrm{s}\right) \propto D_H^3(t_\mathrm{ref};\theta),
\end{align}
where $D_H(t_\mathrm{ref}, \theta)$ is a horizon distance of the
reference detector for a given template at a given timestamp, $t_\mathrm{ref}$.
During an analysis, the \ac{psd} of each detector is measured and
continuously updated. Accordingly, a horizon distance with a fiducial \ac{snr}
of 8 is calculated for every template and stored as a timeseries data.

To compute $P(\vec{O} \mid t_\mathrm{ref}, \mathcal{H}_\mathrm{s})$, we consider the
situation where the \acp{snr} observed by a subset of detectors, $\vec{O}$,
pass a pre-determined threshold, i.e. $\rho>4$, given their horizon distances
at $t=t_\mathrm{ref}$. Recall that we record the horizon distance as a function
of observation time, which makes $t_\mathrm{ref}$ interchangeable with a vector
of the horizon distances for the \textit{operating} detectors, $\vec{D_H}$.  This
\ac{pdf} is computed with sufficient accuracy by simulating isotropically
distributed \ac{gw} sources across a range of distances and performing the
Monte Carlo integration of the detectable sources for a given $\vec{D_H}$ (see
Sec.~III A of \cite{kipp_lr} for more details).

The term $P(\vec{\rho}, \vec{\Delta t}, \vec{\Delta\phi} \mid \vec{O},
t_\mathrm{ref}, \mathcal{H}_\mathrm{s})$ provides a signal-consistency test for
the observed \ac{snr} values and the difference in the event times and phases
among multiple detectors \cite{dtdphi}. Since these parameters defined relative
to those of a reference detector depend on extrinsic parameters, e.g.~the sky
location of a \ac{gw} source, they follow characteristic correlation among
themselves, which helps us distinguish \ac{gw} signals from noise. In this formalism,
we first apply the following coordinate transformation
\begin{align}
    \label{eq:dtdphi_coodtrans}
    \vec{\rho}\rightarrow(\rho_\mathrm{net}, \Delta\vec{\ln \mathcal{D}}),
\end{align}
where $\Delta\vec{\ln \mathcal{D}}$ is a $N-1$ dimensional vector of logarithmic
effective distances for each detector relative to that of a reference detector,
i.e. $\Delta\vec{\ln \mathcal{D}}=\vec{\ln \mathcal{D}} - \ln
\mathcal{D}_\mathrm{ref}$.
Accordingly, to take into account the conversion of volume
elements, we introduce a Jacobian matrix $\boldsymbol{\mathcal{J}}(\vec{\rho})$ in
the calculation of this \ac{pdf}, which reads
\begin{align}
    \label{eq:dtdphi}
    \begin{aligned}
    P(\vec{\rho}, \vec{\Delta t}, \vec{\Delta\phi} \mid \vec{O}, t_\mathrm{ref},\mathcal{H}_\mathrm{s}) &\propto
|\boldsymbol{\mathcal{J}}(\vec{\rho})|\times \rho_\mathrm{net}^{-4}\\
    &\times P(\Delta\vec{\ln \mathcal{D}}, \vec{\Delta t}, \vec{\Delta\phi}\mid \vec{O}, t_\mathrm{ref},
\mathcal{H}_\mathrm{s}).
    \end{aligned}
\end{align}
Here, note that the factor of $\rho_\mathrm{net}^{-4}$, where
$\rho_\mathrm{net}=\sqrt{\sum\rho_i^2}$, stems from the \ac{pdf} of
$\rho_\mathrm{net}$ conditioned on other parameters~\cite{kipp_lr, Schutz_2011},
namely $P(\rho_\mathrm{net}\mid\Delta\vec{\ln \mathcal{D}}, \vec{\Delta t},
\vec{\Delta\phi}, \vec{O}, t_\mathrm{ref}, \mathcal{H}_\mathrm{s})$.
The determinant of the Jacobian matrix, $|\boldsymbol{\mathcal{J}}(\vec{\rho})|$, is given by
\begin{align}
    \label{eq:jacobian_det}
    |\boldsymbol{\mathcal{J}}(\vec{\rho})|=\frac{\rho_\mathrm{net}}{\prod_i \rho_i},
\end{align}
which is derived in \appref{app:jacobian}. 

The PDF $P(\vec{\xi^2} \mid \vec{\rho}, \theta, \mathcal{H}_\mathrm{s})$ gives
the distribution of the $\xi^2$ parameter across the detectors under the
assumption of a GW signal.  Although, in principle, this \ac{pdf} depends on
individual template information $\theta$, for simplicity we approximate that
this $\theta$ dependence is uniform across a group of neighboring templates,
denoted as $\{\bar \theta\}$. We also approximate the $\xi^2$ values given by
multiple detectors to be independent of one another, leading to the
multiplicative form of this joint \ac{pdf}:
\begin{align}
    \label{eq:chisq_signal}
     P\left(\vec{\xi^2} \mid \vec{\rho}, \theta, \mathcal{H}_\mathrm{s}\right) \approx
        \prod_{\mathrm{d}\in\vec{O}} P\left(\xi^2_\mathrm{d}\mid \rho_\mathrm{d}, \{\bar{\theta}\}, \mathcal{H}_\mathrm{s}\right)
\end{align}
such that $\theta\in\{\bar{\theta}\}$. The \ac{pdf} for individual detectors is
approximated as a semi-analytic function. As part of the improvements in the
likelihood ratio for use in \ac{o4}, we have derived this functional form with
improved accuracy as discussed in \secref{sec:chisq_signal_model}.
\subsection{\label{sec:noise_model} Noise model}
In contrast, the \ac{pdf} in the denominator of \eqref{eq:lr} describes how
likely a given event is to be of terrestrial origin. This noise model
\ac{pdf} is factorized as follows
\begin{align}
    \label{eq:noise_model}
    \begin{aligned}
        P(\ldots \mid \mathcal{H}_{\mathrm{n}}) & =P\left(t_\mathrm{ref}, \theta \mid \mathcal{H}_\mathrm{n}\right) \\
        & \times P\left(\vec{O} \mid t_\mathrm{ref}, \theta, \mathcal{H}_\mathrm{n}\right) \\
        & \times P\left(\vec{\Delta t}, \vec{\phi} \mid \vec{O}, \mathcal{H}_\mathrm{n}\right)\\
        & \times P\left(\vec{\rho}, \vec{\xi^2} \mid t_\mathrm{ref}, \theta, \mathcal{H}_\mathrm{n}\right),
    \end{aligned}
\end{align}
where, similarly to \eqref{eq:signal_model}, some of the conditioned parameters
are omitted from each conditional probability when there is no correlation.

The term $P\left(t_\mathrm{ref}, \theta \mid \mathcal{H}_\mathrm{n}\right)$
quantifies the probability of an event with the template $\theta$ being caused by the random noise
fluctuation at $t=t_\mathrm{ref}$. We approximate that the occurrence of such a
noise event obeys a Poisson process and that it is uniform across a template
group $\{\bar{\theta}\}$, finding this \ac{pdf} being proportional to its
occurrence rate\footnote{See \appref{app:trig_rate} for derivation.},
\begin{align}
    \label{eq:p_of_t_noise}
    P\left(t_\mathrm{ref}, \theta \mid \mathcal{H}_\mathrm{n}\right) \propto\mu(t_\mathrm{ref}, \{\bar\theta\}).
\end{align}
Here, $\mu(t_\mathrm{ref}, \{\bar{\theta}\})$ is the mean rate of temporally
coincident events  among \textit{any} combination of operating detectors at
$t=t_\mathrm{ref}$ and for the template group $\{\bar\theta\}$. Therefore,
this is given by adding all the contributions from each subset of detectors, which reads
\begin{align}
    \mu(t_\mathrm{ref}, \{\bar{\theta}\}) = \sum_{\vec{N}\subseteq{\vec{D}}} \mu_{N_1\land N_2\land \ldots},
\end{align}
where $\vec{D}$ represents all the operating detectors, and $\mu_{N_1\land
N_2\land \ldots}$ is the mean rate of coincident events found by
\textit{exactly} the subset $\vec{N}=\{N_1, N_2, \cdots\}$. For example, the two
\ac{ligo} detectors at Hanford ($H$) and Livingston ($L$) are operating,
$\vec{D}=\{H,L\}$, whereas the subsets include $\{H, L\}, \{H\}$ and $\{L\}$.
Note that a single detector case belongs to these subsets, where the
modeled noise event is not necessarily in coincidence. We estimate this rate
from event rates for individual detectors measured at the time segment
associated with $t_\mathrm{ref}$ and for $\{\bar\theta\}$. See Sec.~\Romann{3} D
in \cite{kipp_lr} for more details. From the mean rate of events for each
combination of detectors, it follows that
\begin{align}
    P\left(\vec{O} \mid t_\mathrm{ref}, \theta, \mathcal{H}_\mathrm{n}\right) = \frac{\mu_{O_1\land O_2\land \ldots}}{\mu(t_\mathrm{ref}, \{\bar{\theta}\})}
\end{align}
where we recall that $\vec{O}$ is only the subset of operating detectors that 
detected the event in coincidence.

In the absence of \ac{gw} signals, the coalescence phases and the event time at
each detector's frame relative to $t=t_\mathrm{ref}$ are approximated to be
uniformly distributed, i.e. $P(\vec{\Delta t}, \vec{\phi} \mid
\vec{O}, \mathcal{H}_\mathrm{n})$ is constant.  Also, for $ P(\vec{\rho},
\vec{\xi^2} \mid t_\mathrm{ref}, \theta, \mathcal{H}_\mathrm{n})$, we take an
approximation similar to the signal model and assume the probabilities are
uncorrelated across detectors, so
\begin{align}
    \label{eq:chisq_noise}
    P\left(\vec{\rho}, \vec{\xi^2} \mid t_\mathrm{ref}, \theta, \mathcal{H}_\mathrm{n}\right)\approx\prod_{\mathrm{d}\in\vec{O}} P\left(\rho_\mathrm{d}, \xi^2_\mathrm{d} \mid t_\mathrm{ref}, \{\bar\theta\}, \mathcal{H}_\mathrm{n}\right).
\end{align}
Unlike \eqref{eq:chisq_signal} in the signal model, we construct this \ac{pdf}
in \eqref{eq:chisq_noise} for individual detectors by collecting non-coincident
events (i.e. found by only one detector) during observation and histogram them
in the $(\rho_\mathrm{d}, \xi^2_\mathrm{d})$ parameter space. When calculating
the ranking statistic for given $(\vec{\rho}, \vec{\xi^2})$ values,
\eqref{eq:chisq_noise} is evaluated after a smoothing process is applied on each
$(\rho_\mathrm{d}, \xi^2_\mathrm{d})$ histogram in order to obtain a smooth
distribution of the ranking statistic $\mathcal{L}$.  We note that this \ac{pdf}
depends on the event time $t_\mathrm{ref}$ as \gstlal{} collects $(\vec{\rho},
\vec{\xi^2})$ values continuously during its analysis. This allows us to track
the noise characteristics evolving over time and provide a more accurate
estimate of $\mathcal{L}$. Technically, we model this time-dependent \ac{pdf} in
a different manner between the online and offline analysis, e.g. cumulatively
for online analysis as compared to segment-wise for offline analysis.

\subsection{Single-detector events}
So far, the signal (noise) model formulated in \eqref{eq:signal_model}
(\eqref{eq:noise_model}) implicitly assumes an event identified in coincidence
across multiple detectors. Yet, it is possible that real \ac{gw} signals trigger
only one detector for various reasons, e.g. only one detector operates at the
time of such an event in the first place or the other detectors observe the
event below the \ac{snr} threshold due to their blind spots. To handle these
cases, \gstlal{} has been capable of evaluating the likelihood ratio for
single-detector events since \ac{o2}~\cite{gstlal_o2}. In this situation, all
the vector quantities shown in \eqref{eq:lr}, e.g. $\vec{t}, \vec{\phi},
\vec{\rho}$ and $\vec{\xi^2}$, are considered as scalar quantities. In particular,
\eqref{eq:dtdphi} is simplified as
\begin{align}
    \label{eq:dtdphi_1ifo}
    P(\rho\mid \vec{O}, t_\mathrm{ref},\mathcal{H}_\mathrm{s}) \propto\rho^{-4},
\end{align}
and the multiplicative \acp{pdf} shown in
\eqsref{eq:chisq_signal}{eq:chisq_noise} reduces to a function with single term.

Also, since \ac{o2} we have introduced a penalty term in the likelihood ratio to
properly downrank the statistical significance of such events. While this is a
somewhat arbitrary parameter to tune, we adopt the penalty value of $-13$ in log
likelihood ratio after assessing the distribution of single-detector events of
terrestrial origin we detected during the Mock Data Challenge conducted as a
preparation of \ac{o4}. See~\cite{o4_performance} for more details.
\section{Developments}
\label{sec:dev}
Having described the components of the numerator and denominator in the 
likelihood ratio in Sec.~\ref{sec:lr}, in this section we describe the improvements 
to the likelihood ratio made in advance of \ac{o4}.

\subsection{Removal of signal contamination}
\label{sec:signal_contamination}

As described in \secref{sec:noise_model}, we construct the \ac{pdf}s given in
\eqref{eq:chisq_noise} by forming histograms of non-coincident events in $(\rho_\mathrm{d},
\xi^2_\mathrm{d})$ space.  Given the noise hypothesis, we only collect events originating
from noise to populate these histograms. Since we expect that
\ac{gw} signals coincide between detectors, most of the events originating from
\ac{gw} signals are properly eliminated by requiring the events to be
non-coincident while more than one detector is running.

However, it is possible that a \ac{gw} signal is recovered as a non-coincident
event when more than one detector is running, due to both astrophysical and
terrestrial reasons.  An example of an astrophysical reason is the \ac{gw} signal
originating from a sky location detectable for only one detector, whereas a
terrestrial reason would be that only one detector is sensitive enough to detect
the signal. In addition, a loud \ac{gw} signal might be recovered as a
coincident event in one template group with good match, but as a non-coincident
event in some neighboring groups with less match, which populates the background
histograms of those groups.  This can potentially cause the \ac{pdf}s in
\eqref{eq:chisq_noise} to be constructed incorrectly, which is commonly called
\textit{signal contamination} of the $\rho - \xi^2 $ histograms, leading to a biased
significance estimation.

\begin{figure*}[htbp]
\includegraphics[width=\textwidth]{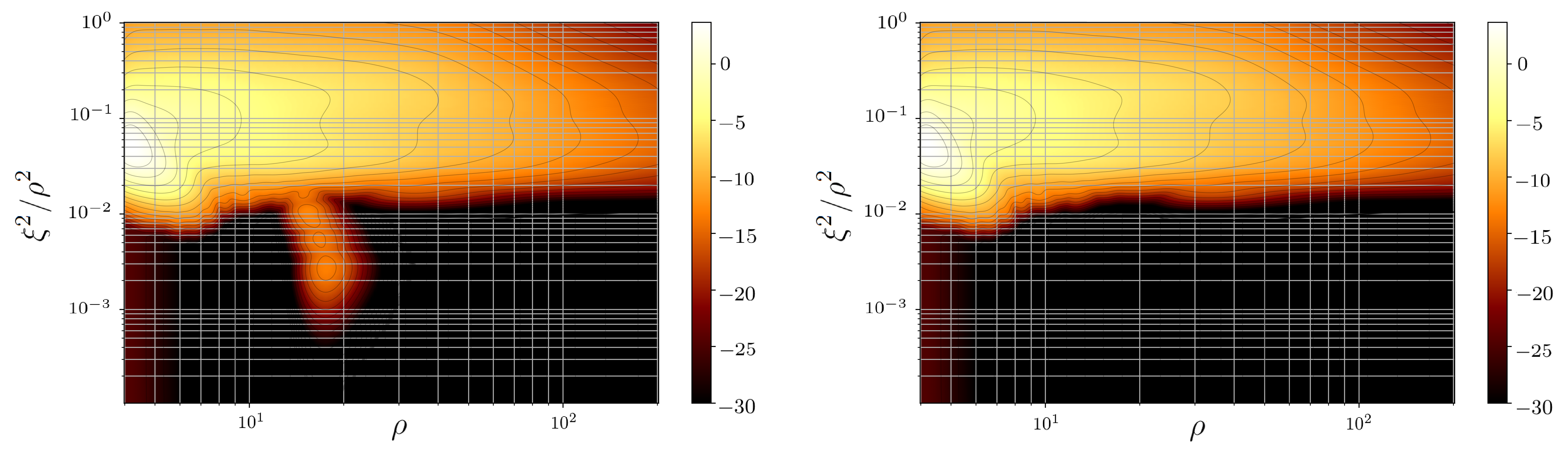}
\caption{\label{fig:signal_contamination}
Left: An example of signal contamination in a $\rho - \xi^2$ \ac{pdf} for
Livingston.  Right: The same \ac{pdf}, with the contamination removed using the
mechanism described in \secref{sec:signal_contamination}. The color scheme
encodes the probability density in a logarithmic scale with a brighter region
having larger values. Note that kernel smoothing has been applied to these
\ac{pdf}s.
}
\end{figure*}

To prevent this, it is necessary to remove the events associated with such
\ac{gw} signals from the background histograms. However, there exists a tradeoff
regarding which events one should choose to remove.  Blindly removing all
potential contamination risks undesirably eliminating actual noise events with
high significance.  Therefore, this contamination removal is designed to be a
manual operation so that the user can decide the criterion by which to remove
background events.  For the \gstlal{} analysis, while the data is being analyzed
and the histograms are being populated, we keep a temporary record of the non-coincident
events from the last \SI{5000}{\second}. If an event is above the open public
alert significance threshold~\cite{opa}, it is considered as a potential
contamination, and events within a \SI{10}{\second} window around the event of interest
are permanently stored across all the template bins. Subsequently, these events are subtracted from the
histograms so that the likelihood ratio can be evaluated without their contamination.

An example of the effect of signal contamination on a $\rho - \xi^2$ \ac{pdf},
and the subsequent removal of the contamination are shown in
\figref{fig:signal_contamination}.  More information about the mechanism of
removing signal contamination and its effect on the sensitivity of the \gstlal{}
search can be found in~\cite{count_tracker}.

\subsection{\label{sec:combochisq} Incorporation of bank $\xi^2$}
Although \ac{snr} is known to be the optimal ranking statistic under Gaussian
noise, nonstationary noise deviating from the Gaussian distribution, often
referred to as \textit{glitches}, can yield large \ac{snr} values and mimic a
\ac{gw} signal. Therefore, \ac{snr} alone does not characterize \ac{gw} signals
sufficiently. To deal with this, \gstlal{} adopts signal-based vetoes by
introducing the $\xi^2$ parameter.  Conceptually, these vetoes perform a
consistency test between \ac{snr} values of a given event across a parameter
space of our interest and its expected evolution in the presence of a signal.
Up to \ac{o3}, \gstlal{} calculated the $\xi^2_\mathrm{a}$
parameter\footnote{Note that a subscript $_\mathrm{a}$ indicates the parameter
derived from an auto-correlation function to make a distinction from
\textit{bank}-$\xi^2$ parameter we mention subsequently.} by considering
residual between \ac{snr} time series of a given event associated with a
template $\alpha$ and its auto-correlation function.  In what follows, we start
with our conventional formalism of the $\xi^2_\mathrm{a}$ parameter and
subsequently introduce an alternative $\xi^2$ parameters, which we call
\textit{bank-}$\xi^2$, developed for \ac{o4}.

Given the two polarizations of \acp{gw}, we conventionally describe a template
as complex values whose real and imaginary parts represent either $+$ or
$\times$ polarization so that
\begin{align}
    {h}_\alpha[\tau] = {h}_\alpha^+[\tau] + i{h}_\alpha^\times[\tau].
\end{align}
With this representation of a template, we express the \ac{snr} timeseries
associated with the template $\alpha$ as the following complex values
\begin{align}
    \hat{z}_\alpha[t] =  \hat{z}_\alpha^+[t] +i \hat{z}_\alpha^\times[t].
\end{align}
Note that for the rest of this paper, we denote a hat symbol $\hat{x}$ to
explicitly indicate $x$ is a random variable subject to noise fluctuation. The
real and imaginary parts are the matched filter output for the template with
either polarization, i.e.
\begin{align}
    \hat{z}_\alpha^A[t] = \sum_\tau {h}_\alpha^A[\tau]\hat{s}[t+\tau].
\end{align}
Here $\hat{s}[t]$ is whitened strain timeseries given from the upstream of a
pipeline and ${h}_\alpha^A[\tau]$ is a whitened template normalized such that
\begin{align}
    \label{eq:temp_norm}
    \sum_\tau {h}_\alpha^A[\tau]{h}_\alpha^A[\tau]=1
\end{align}
for the two polarizations $A={\{+,\times\}}$.

Being analogous to the $\chi^2$ test in statistics, the
$\hat{\xi}^2_\mathrm{a}$ parameter is defined as~\cite{gstlal_cody}
\begin{align}
    \label{eq:autochisq_def}
    \hat{\xi}^2_\mathrm{a} = \frac{1}{N_\mathrm{a}}\sum_t\left|\hat{z}_\alpha[t] - \hat{z}_\alpha[0]R_\alpha[t]\right|^2,
\end{align}
where $R_\alpha[t]$ is a complex auto-correlation function of the whitened template
$\alpha$
\begin{align}
    R_\alpha[t] = \frac{1}{2}\sum_\tau {h}_\alpha[\tau]{h}_\alpha^*[t+\tau],
\end{align}
normalized as $R_\alpha[0]=1$. $N_\mathrm{a}$ is
given by
\begin{align}
    N_\mathrm{a} = \sum_t\left\{2 - 2\left|R_\alpha[t]\right|^2\right\}
\end{align}
so that the expectation value of $\hat{\xi}^2_\mathrm{a}$ is unity, i.e.
$\langle\hat{\xi}^2_\mathrm{a}\rangle=1$. Also, $t=0$ in
\eqref{eq:autochisq_def} is chosen from the peak of
$\left|\hat{z}_\alpha[t]\right|^2$, which is the best estimate of the merger
time of a \ac{cbc} signal. In practice, the summation over $\tau$ does not cover
an infinite range, but rather one needs to select a finite auto-correlation
length. For \ac{o4}, we adopt the length of 701 (351) samples for templates with
the chirp mass below (above) 15$M_\odot$, respectively.  Note that in the case
of the noiseless limit and a signal being identical to the associated template,
the evolution of $\hat{z}_\alpha[t]$ follows the auto-correlation function with
the proper scaling and phase shift accounted. Thus under Gaussian noise, the
residual elements, $\hat{z}_\alpha[t] - \hat{z}_\alpha[0]R_\alpha[t]$, are each
distributed as a Gaussian with zero mean and some correlation across samples at
different timestamps. Given this property, the $\hat{\xi}^2_\mathrm{a}$
parameter allows us to assess the consistency of the \ac{snr} evolution over
time by comparing the $\hat{\xi}^2_\mathrm{a}$ value against its expected
distribution under the signal and noise models.

In general, $\xi^2$ parameters can be defined differently to perform a
consistency test of \ac{snr} values in a parameter space other than time.
Hence, we have developed \textit{bank-}$\xi^2$ parameter~\cite{chad_thesis},
denoted as $\hat{\xi}^2_\mathrm{b}$, to explore the \ac{snr} evolution in
template-bank space. Similarly to the conventional $\hat{\xi}^2_\mathrm{a}$
parameter shown in \eqref{eq:autochisq_def}, we evaluate this statistic for
an event associated with the template $\alpha$ by calculating the \ac{snr}
residual across a group of templates denoted as $\{\bar{\theta}\}$:
\begin{align}
    \label{eq:bankchisq_def}
    \hat{\xi}^2_\mathrm{b} = \frac{1}{N_\mathrm{b}}\sum_{\beta\in\{\bar{\theta}\}}\left|\hat{z}_\beta[0] - \frac{1}{2}(h_\beta | h_\alpha)\hat{z}_\alpha[0]\right|^2,
\end{align}
where $(h_\beta | h_\alpha)$ is the matched filter output between the two
\textit{complex} whitened templates, i.e.
\begin{align}
    (h_\beta | h_\alpha) = \sum_\tau {h}_\beta[\tau]{h}_\alpha^*[\tau].
\end{align}
Here $N_\mathrm{b}$ is the normalization factor 
\begin{align}
    N_\mathrm{b} = \sum_{\beta\in\{\bar{\theta}\}}\left\{2 - \frac{1}{2}\left|(h_\beta | h_\alpha)\right|^2\right\}
\end{align}
so that the expectation value of $\hat{\xi}_\mathrm{b}^2$ is unity, i.e.
$\langle\hat{\xi}_\mathrm{b}^2\rangle=1$. Given the different parameter space which
the $\hat{\xi}^2_\mathrm{b}$ parameter involves, its consistency test provides additional
information about \ac{gw} signals complementary to the $\hat{\xi}^2_\mathrm{a}$
parameter.

To take advantage of the benefits from both $\hat{\xi}^2_\mathrm{a}$ and
$\hat{\xi}^2_\mathrm{b}$, we introduce a $\hat{\xi}^2_\mathrm{ab}$ parameter that combines the
residual components of $\hat{\xi}^2_\mathrm{a}$ and $\hat{\xi}^2_\mathrm{b}$ as
follows
\begin{align}
    \hat{\xi}^2_\mathrm{ab} =\frac{1}{N_\mathrm{a} + N_\mathrm{b}}&\left(N_\mathrm{a}\hat{\xi}^2_\mathrm{a} + N_\mathrm{b}\hat{\xi}^2_\mathrm{b}\right)\\
    \begin{split}
    = \frac{1}{N_\mathrm{a} + N_\mathrm{b}}&\left\{\sum_t\left|\hat{z}_\alpha[t] - \hat{z}_\alpha[0]R_\alpha[t]\right|^2\right.\\
    &\left.+\sum_{\beta\in\{\bar{\theta}\}}\left|\hat{z}_\beta[0] - \frac{1}{2}(h_\beta | h_\alpha)\hat{z}_\alpha[0]\right|^2\right\},
    \end{split}
    \label{eq:combinedchisq_def}
\end{align}
where the normalization factor is applied similarly such that
$\langle\hat{\xi}^2_\mathrm{ab}\rangle=1$. In \secref{sec:results}, we show the
sensitivity improvement of \gstlal{} pipeline due to this
$\hat{\xi}^2_\mathrm{ab}$ parameter, using a simulated search for injected
\ac{gw} signals.

\subsection{Upgraded \label{sec:chisq_signal_model}$\rho - \xi^2$ signal model}
Previously in \ac{o3}, $P(\vec{\xi^2} \mid \vec{\rho}, \theta, \mathcal{H}_\mathrm{s})$
in \eqref{eq:signal_model} followed an analytic function that was empirically
obtained by tuning free parameters. Here, we describe a more accurate
approximation of this \ac{pdf} derived from the statistical properties of the
$\xi^2$ parameter. In the rest of this subsection, we assume the signal
hypothesis where a \ac{gw} signal is present in Gaussian noise unless stated
otherwise.
\subsubsection{General formalism}
We start with
the general definition of the $\xi^2$ parameter:
\begin{align}
    \label{eq:generalchisq_def}
    \hat{\xi}^2= \frac{1}{N}\sum_{j}\left|\hat{z}_j - C_{ij}\hat{z}_i\right|^2,
\end{align}
where $\hat{z}_i$ is a matched filter output at a reference point, denoted by the
index $i$, of an arbitrary parameter and $N$ is a normalization factor, which reads
\begin{align}
    N = \sum_{j}\left\langle\left|\hat{z}_j - C_{ij}\hat{z}_i\right|^2\right\rangle.
\end{align}
$C_{ij}$ is a coefficient to construct expected morphology of $\{\hat{z}_j\}$,
which can be thought of as a transfer function that relates $\hat{z}_i$ to
$\hat{z}_j$, and this is given by $C_{ij}=\langle \hat{z}_j\hat{z}_i^*\rangle$
in the absence of a signal.
% for example, $C_{ij}=R(t_j - t_i)$, where $t_i=0$.
Although $\xi^2, N$ and quantities derived from these are specific to
the reference point, indexed by $i$, of the parameter of interest, for brevity
we do not explicitly indicate its dependence in these symbols.

Here, introducing a vector of the residual components
\begin{align}
    \vec{U} =\{\hat{U}_j\}= \hat{z}_j - C_{ij}\hat{z}_i,
\end{align}
it follows that
$\hat{\xi}^2\propto \vec{U}^\dag\vec{U}$, where $^\dag$ represents Hermitian
transpose of a given vector or matrix. Given the statistical properties of the matched
filter output $\hat{z}_i$, $\vec{U}$ obeys a complex Gaussian distribution. In
practice, there remains a systematic mismatch between the true signal and its
associated template, leading to a non-zero mean of $\vec{U}$, which we denote as
$\vec{\mu}=\langle \vec{U}\rangle$. Also, its covariance matrix is given by
\begin{align}
    \label{eq:sigma_def}
    \Sigma_{jk} &= \langle (\hat{U}_j^* - \mu_j^*)(\hat{U}_k - \mu_k)\rangle\\
        &=C_{jk} - C_{ij}^*C_{ik}.
\end{align}
Consequently, one can find that
\begin{align}
    \label{eq:Usq}
    \vec{U}^\dag\vec{U} &= \sum_j\lambda_j|\hat{n}_j + \eta_j|^2,
\end{align}
where $\{\lambda_j\}$ is an eigenvalue of the covariance matrix
$\boldsymbol{\Sigma}$, $\{\hat{n}_j\}$ is normal random variables, and
$\{\eta_j\}$ is the $j$-th component of $\vec{\mu}$ projected onto the
eigenvector space of $\boldsymbol{\Sigma}$. See \appref{app:chisq_derive} for
the derivation of \eqref{eq:Usq}. 

The expression in RHS of \eqref{eq:Usq} is known as \textit{generized chi-square
distribution} and there does not exist any closed-form expression of this
\ac{pdf}.  However, this can be well approximated with a \textit{single}
chi-square distribution\footnote{Eq.(13) of \cite{nasa_chisq} shows the
approximation by a Gamma distribution, but this is equivalent to a chi-square
distribution with the proper degree of freedom ($n_\mathrm{eff}$) and multiplied
factor in the coordinate ($b$).} whose mean and variance are matched up with
those derived from \eqref{eq:Usq}~\cite{nasa_chisq}, which yields
\begin{align}
    \label{eq:xi_mean}
    \langle\hat{\xi}^2\rangle &= \frac{1}{N}\left\{\sum_j\lambda_j + \sum_j\lambda_j|\eta_j|^2\right\},\\
     &= \frac{1}{N}\left\{\mathrm{Tr}\boldsymbol{\Sigma} + \vec{\mu}^\dag\vec{\mu}\right\},\\
    \label{eq:xi_var}
    \mathrm{Var}(\hat{\xi}^2) &= \frac{2}{N^2}\left\{\sum_j\lambda_j^2 + 2\sum_j\lambda_j^2|\eta_j|^2\right\}.\\
     &= \frac{2}{N^2}\left\{\mathrm{Tr}\boldsymbol{\Sigma}^2 + 2\vec{\mu}^\dag\boldsymbol{\Sigma}\vec{\mu}\right\}.
\end{align}
Given these known mean and variance, we parametrize the desired chi-square
distribution with the following two parameters:
\begin{align}
    \label{eq:n_eff}
    n_\mathrm{eff} = \frac{2\langle\hat{\xi}^2\rangle^2}{\mathrm{Var}(\hat{\xi}^2)}\\
    \label{eq:bias}
    b = \frac{\mathrm{Var}(\hat{\xi}^2)}{2\langle\hat{\xi}^2\rangle},
\end{align}
such that $\chi^2(x/b; n_\mathrm{eff})$ reproduces its mean and
variance\footnote{$\chi^2(x; n)$ represents a chi-square \ac{pdf} with $n$
degrees of freedom as a function of $x$.} consistent with
\eqsref{eq:xi_mean}{eq:xi_var}, respectively. Therefore, we approximate that
$P(\xi^2 \mid \rho, \mathcal{H}_\mathrm{s})\approx\chi^2(\xi^2/b;
n_\mathrm{eff})$. Note that this chi-square distribution depends on $\rho$
implicitly through $\{\eta_j\}$, which will be described in more details below
in the case of the $\xi_\mathrm{a}^2$ parameter.
\subsubsection{Application to $\hat{\xi}^2_\mathrm{a}$ parameter}
\label{sec:chisq_signal_model_app}
We apply the above formalism to \gstlal's conventional $\hat{\xi}^2_\mathrm{a}$
parameter based on its definition shown in \eqref{eq:autochisq_def}.  In this
case, $i,j$ indices in \eqref{eq:generalchisq_def} represent timestamps of \ac{snr}
timeseries and specifically $t_i=0$. Also, from $C_{ij}=R[t_j-t_i]$ and
\eqref{eq:sigma_def}, it follows that
\begin{align}
    \label{eq:autochisq_covmat}
    \Sigma_{jk} =R[t_k-t_j] - R^*[t_j]R[t_k].
\end{align}
Regarding the mismatch between a true signal and its associated template, for
simplicity we approximate it with the original auto-correlation function $R[t]$
overall scaled with the detected \ac{snr} and a fractional mismatch factor $k$,
such that 
\begin{align}
    \label{eq:mu_approx}
    \mu[t]=\langle \hat{z}[t] - \hat{z}[0]R[t]\rangle\approx k\rho R[t].
\end{align}
It is nontrivial to assess the accuracy of this approximation. Another possible
approach would be to take the cross-correlation between a neighboring pair of
templates. We leave the investigation of this and other avenues for improving
the accuracy of this approximation to future work.

Nevertheless, \eqref{eq:mu_approx} allows for rewriting \eqsref{eq:xi_mean}{eq:xi_var} in terms
of $\vec{R}=\{R[t]\}$ as follows
\begin{align}
    \label{eq:xi_mean_R}
    \langle\hat{\xi}^2_\mathrm{a}\rangle &= \frac{1}{N}\left\{\mathrm{Tr}\boldsymbol{\Sigma} + \rho^2k^2\vec{R}^\dag\vec{R}\right\},\\
    \label{eq:xi_var_R}
    \mathrm{Var}(\hat{\xi}^2_\mathrm{a}) &= \frac{2}{N^2}\left\{\mathrm{Tr}\boldsymbol{\Sigma}^2 + 2\rho^2k^2\vec{R}^\dag\boldsymbol{\Sigma}\vec{R}\right\}.
\end{align}
This implies that, for given $\boldsymbol{\Sigma}$, $\rho$ and $k$, a single
chi-square distribution $\chi^2(\xi^2/b; n_\mathrm{eff})$ is constructed based
on \eqsref{eq:n_eff}{eq:bias}. Furthermore, this \ac{pdf} is marginalized over a
range of $k$, \SIrange{0.1}{30}{\percent}, as currently implemented in \gstlal.
We iterate this process across different $\rho$ values until the $(\rho,\xi^2)$
parameter space is sufficiently covered. See \figref{fig:sig_scatter} for a
visualization.

\eqref{eq:autochisq_def} shows that the $\hat{\xi}_\mathrm{a}^2$ parameter
and its \ac{pdf} are specific to the whitened auto-correlation function
$R_\alpha[t]$ of a particular template $\alpha$, which in turn implicitly
depends on a \ac{gw} detector through whitening by its \ac{psd}. Also, as indicated by
\eqref{eq:chisq_signal}, for the sake of memory management we assign a common
$P(\vec{\xi^2} \mid \vec{\rho}, \theta, \mathcal{H}_\mathrm{s})$ \ac{pdf} to a
group of neighboring templates $\{\bar{\theta}\}$, which contains around 1000
templates in average, assuming that \ac{gw} events associated with any of these
templates follow the same signal model.  To construct this representative
\ac{pdf}, the median of $\langle\hat{\xi}^2_\mathrm{a}\rangle$ and
$\mathrm{Var}(\hat{\xi}^2_\mathrm{a})$ values are computed for each template
group. The validity of this \ac{pdf} depends on the similarity among the
templates within each group, i.e. the efficiency of template grouping, whose
details are described in \cite{shio_template_bank}. As a result, we construct
one $ P\left(\xi^2_\mathrm{d} \mid \rho_\mathrm{d}, \{\bar{\theta}\},
\mathcal{H}_\mathrm{s}\right)$ signal model per template group per \ac{gw}
detector, and for a given detected event, $P(\vec{\xi^2} \mid \vec{\rho},
\theta, \mathcal{H}_\mathrm{s})$ is evaluated in a multiplicative form with
regard to participating \ac{gw} detectors $\{\vec{O}\}$ as shown in
\eqref{eq:chisq_signal}.

\subsection{KAGRA integration}
KAGRA participated in the joint observation with GEO600~\cite{geo600} detector
at the end of \ac{o3}~\cite{o3gk}, and is planning to join the full \ac{gw}
detector network including Advanced \ac{ligo} and Virgo detectors during
\ac{o4}~\cite{obssci}. Accordingly, \ac{gw} detection pipelines need to
incorporate the additional detector and conduct analysis across all the
detectors in coincidence. With regard to \gstlal's likelihood ratio, the two
\acp{pdf} in the signal model, e.g. $P(\vec{O} \mid t_\mathrm{ref},
\mathcal{H}_\mathrm{s})$ and $P(\Delta\vec{\ln \mathcal{D}}, \vec{\Delta t},
\vec{\Delta\phi}\mid \vec{O}, t_\mathrm{ref}, \mathcal{H}_\mathrm{s})$ are
relevant to this integration.  Here, we illustrate these \acp{pdf} in the
presence of KAGRA and briefly discuss its characteristics.

\begin{figure}[t]
    \includegraphics[width=\linewidth]{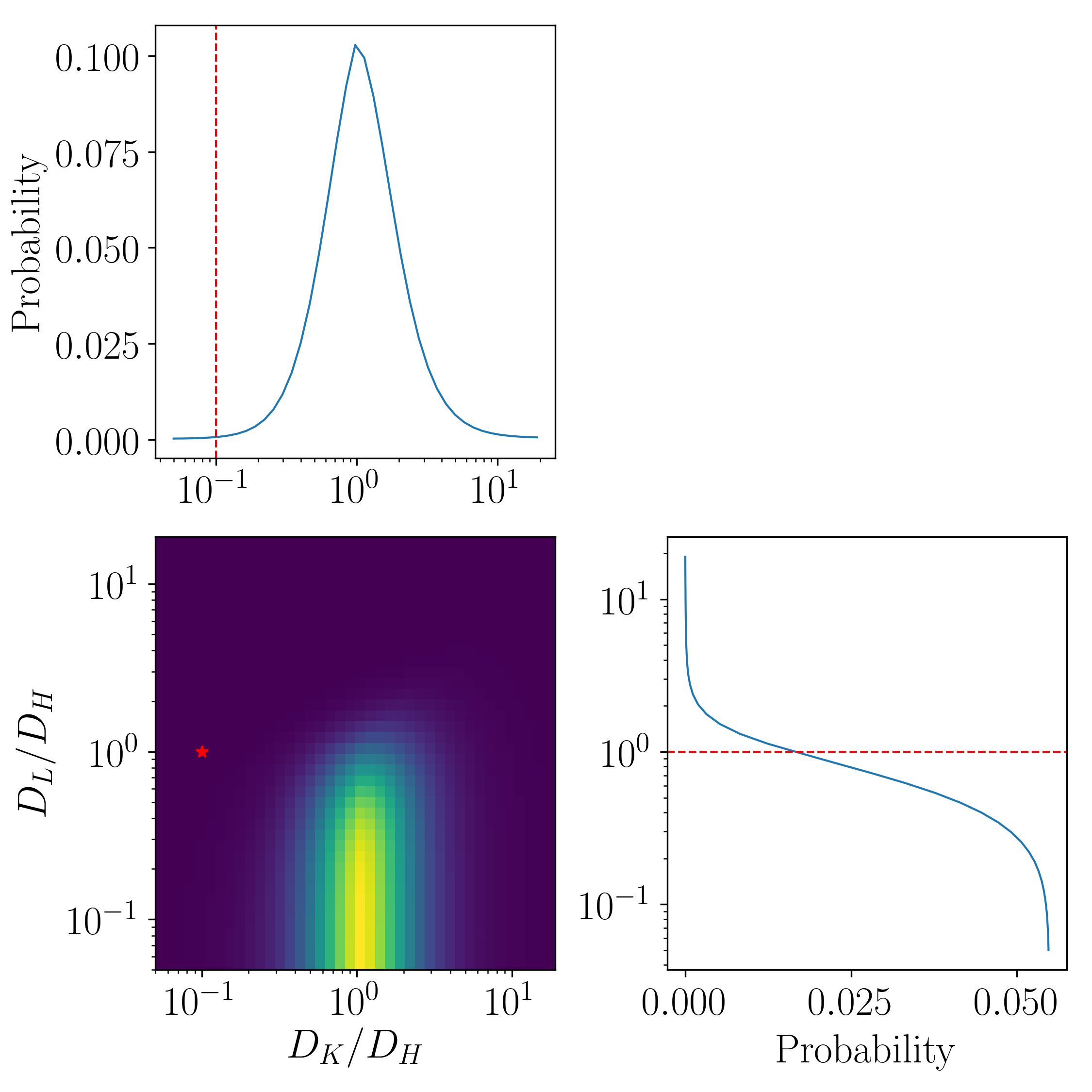}
    \caption{\label{fig:p_ifo_hk}
    Example $P(\{H, K\} \mid t_\mathrm{ref},\mathcal{H}_\mathrm{s})$ while
    \ac{ligo} Hanford, Livingston and KAGRA operating, is evaluated as a
    function of fractional horizon distance between KAGRA ($D_K$) and \ac{ligo}
    Hanford ($D_H$) or between Hanford and Livingston ($D_L$), respectively.
    The color scheme in the two-dimensional plot encodes the probability density
    with a brighter region having larger values.
    The red marker indicates the reasonable values of the fractional horizon
    distance among these detectors during \ac{o4}, being far from the peak.
    }
\end{figure}

\figref{fig:p_ifo_hk} shows, as an example, the probability of only \ac{ligo}
Hanford and KAGRA forming a coincident event while the two \ac{ligo} detectors
and KAGRA operate, i.e. $P(\{H, K\} \mid
t_\mathrm{ref},\mathcal{H}_\mathrm{s})$. This probability is evaluated as a
function of horizon distances of KAGRA ($D_K$) and \ac{ligo} Livingston ($D_L$)
relative to that of \ac{ligo} Hanford ($D_H$). Recall that $t_\mathrm{ref}$ is
interchangeable with a vector of horizon distances, and hence taking different
$t_\mathrm{ref}$ values is equivalent to exploring the two-dimensional parameter
space $(D_K/D_H, D_L/D_H)$. The peak of the probability is located at
$D_K/D_H\sim1$ and $D_L/D_H\ll 1$, which is expected for both \ac{ligo} Hanford
and KAGRA detectors to observe a signal and for \ac{ligo} Livingston detector to
miss it. In practice, however, reasonable values of the fractional horizon
distance among these detectors during \ac{o4} is far from the peak as indicated
by the red marker in \figref{fig:p_ifo_hk}, implying heavy downranking if
such an event is observed.

\begin{figure}[t]
    \includegraphics[width=\linewidth]{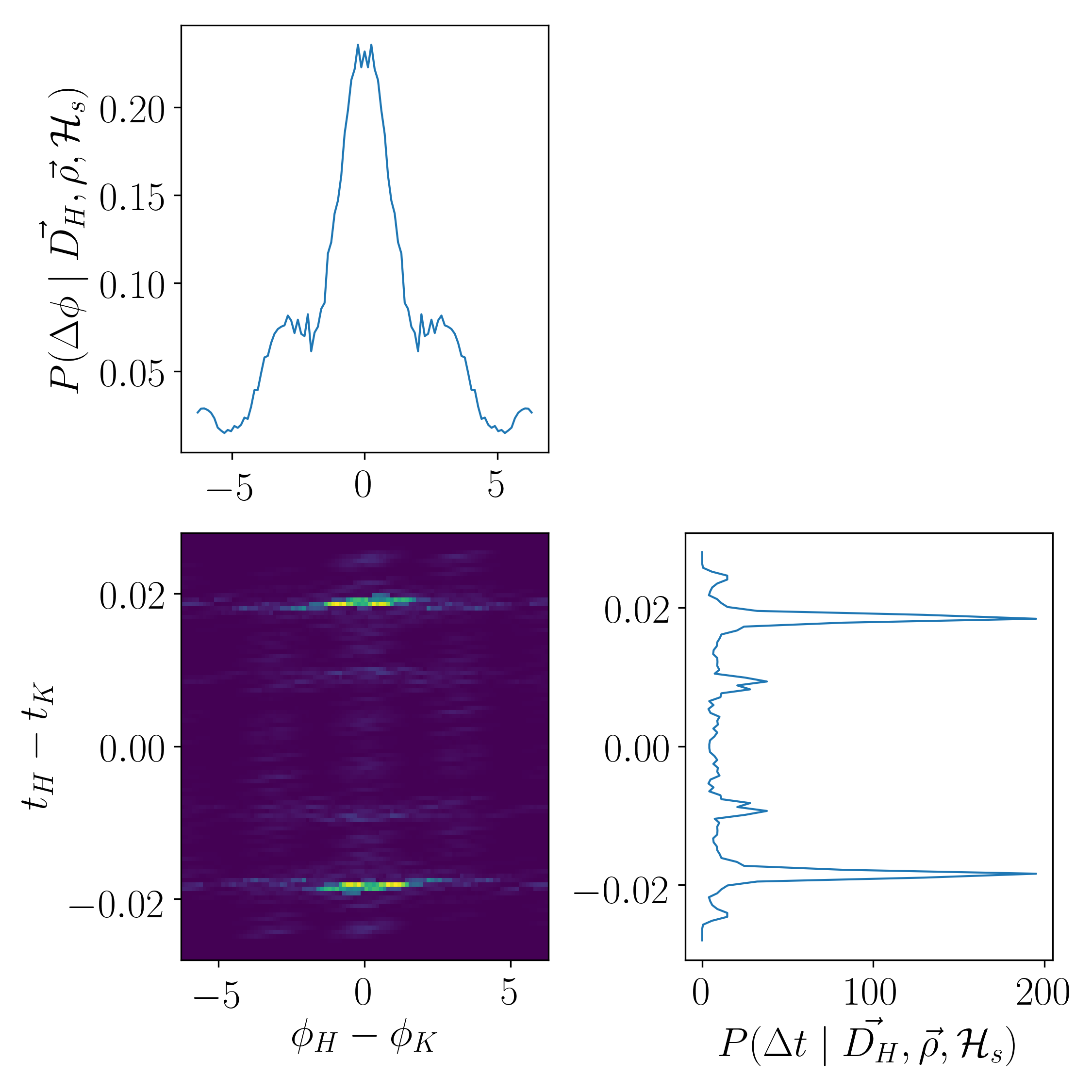}
    \caption{\label{fig:dtdphi_hk}
    Example $P(\vec{\Delta{t}}, \vec{\Delta{\phi}}\mid \{H, K\}, t_\mathrm{ref},
    \mathcal{H}_\mathrm{s})$ and its marginalized \ac{pdf} for LIGO Hanford and
    KAGRA.  The color scheme in the two-dimensional plot encodes the probability
    with a brighter region having larger values.  Here we set the
    horizon distances of LIGO Hanford and KAGRA to be 410 Mpc and 6 Mpc
    respectively to be consistent with the projected \ac{o4} \ac{psd} of the two
    detectors~\cite{obssci}, and show a slice of the three-dimensional \ac{pdf}
    where the ratio of the effective distance between the two detectors is
    somewhat reasonable given their horizon distances.
    }
\end{figure}

In \figref{fig:dtdphi_hk} we illustrate the two-dimensional \ac{pdf} for the
difference in \ac{gw} arrival time and phase between LIGO Hanford and KAGRA,
i.e. $P(\vec{\Delta{t}}, \vec{\Delta{\phi}}\mid \{H, K\}, t_\mathrm{ref},
\mathcal{H}_\mathrm{s})$. Here we set the horizon distances\footnote{One can
obtain the \ac{bns} range, which is commonly used as a measure of detector
sensitivity in literature, by multiplying the horizon distance by the orientation-average factor
$\mathcal{F}\simeq(2.2627)^{-1}$~\cite{findchirp}.} of LIGO Hanford and KAGRA for a typical
\ac{bns} source to be 410 Mpc and 6 Mpc, being consistent with the projected
\ac{o4} \ac{psd} of the two detectors~\cite{obssci}, and their representative
\acp{snr} to be 5 and 4, respectively. Therefore, \figref{fig:dtdphi_hk}
represents a slice of the three-dimensional \ac{pdf} $P(\Delta\vec{\ln
\mathcal{D}}, \vec{\Delta t}, \vec{\Delta\phi}\mid \vec{O}, t_\mathrm{ref})$
where the effective distance ratio between the two detectors is given by their
horizon distances and representative \acp{snr} mentioned above.  Also, we
observe the characteristic structure in the marginalized $\Delta t$
distribution, $P(\Delta t\mid \{H, K\}, t_\mathrm{ref},
\mathcal{H}_\mathrm{s})$. This can be explained by the extreme value of the
effective distance ratio between the two detectors,
$\mathcal{D}_H/\mathcal{D}_K=16.5$, which allows only a narrow range of
extrinstic parameters to contribute to the \ac{pdf}. In particular, such a
limited range of sky position is manifested by the noticeable structure in
$P(\Delta t\mid  \{H, K\}, t_\mathrm{ref}, \mathcal{H}_\mathrm{s})$.
\section{Results}
\label{sec:results}
\subsection{Sensitive space-time volume}
\label{sec:vt}
We conduct an injection study using \gstlal{} with the new features described in
\secref{sec:dev} and discuss the improvement in the pipeline's sensitivity,
focusing on the bank-$\xi^2$ incorporation and the upgraded $\rho-\xi^2$ signal
model. For each injection run, to quantify the sensitivity we measure the
\ac{vt} as a function of \ac{far}, which is defined as
\begin{align}
    \label{eq:vt}
    VT(\mathrm{FAR}) = T\int_{0}^{\infty}\epsilon(z, \mathrm{FAR})\frac{d V_c(z)}{dz}\frac{1}{1+z}\diff z,
\end{align}
where $T$ is the duration of a simulated observation, $\epsilon(z,
\mathrm{FAR})$ is the detection efficiency for the \ac{gw} signals which are
injected at the redshift in [$z, z+\mathrm{d} z$] and recovered at \ac{far}
below a given threshold, and $V_c(z)$ is the comoving volume at the redshift of
$z$. Note that \ac{vt} depends on the source distribution of injected \ac{gw}
signals, and in what follows, we apply the same injection set between two runs
for valid \ac{vt} comparison.

\subsection{\label{sec:results_combinedchisq} Incorporation of bank $\xi^2$}
For a simulated observation, we analyze the \ac{o3} dataset between 18 April
2019 16:46 UTC and 26 April 2019 17:14 UTC with 86606 synthetic \ac{cbc} signals
injected. This injection set contains signals from \acp{bns} whose component
masses ($m_1, m_2$) go up to $3M_\odot$, \acp{bbh} whose $m_1,m_2$ go up to
$50M_\odot$, and \acp{imbh} whose $m_1, m_2$ go up to $300M_\odot$. For this
study, we use the template bank used for \ac{o3}~\cite{gwtc-3}, which covers the
entire set of injections.
\begin{figure}[t]
    \includegraphics[width=\linewidth]{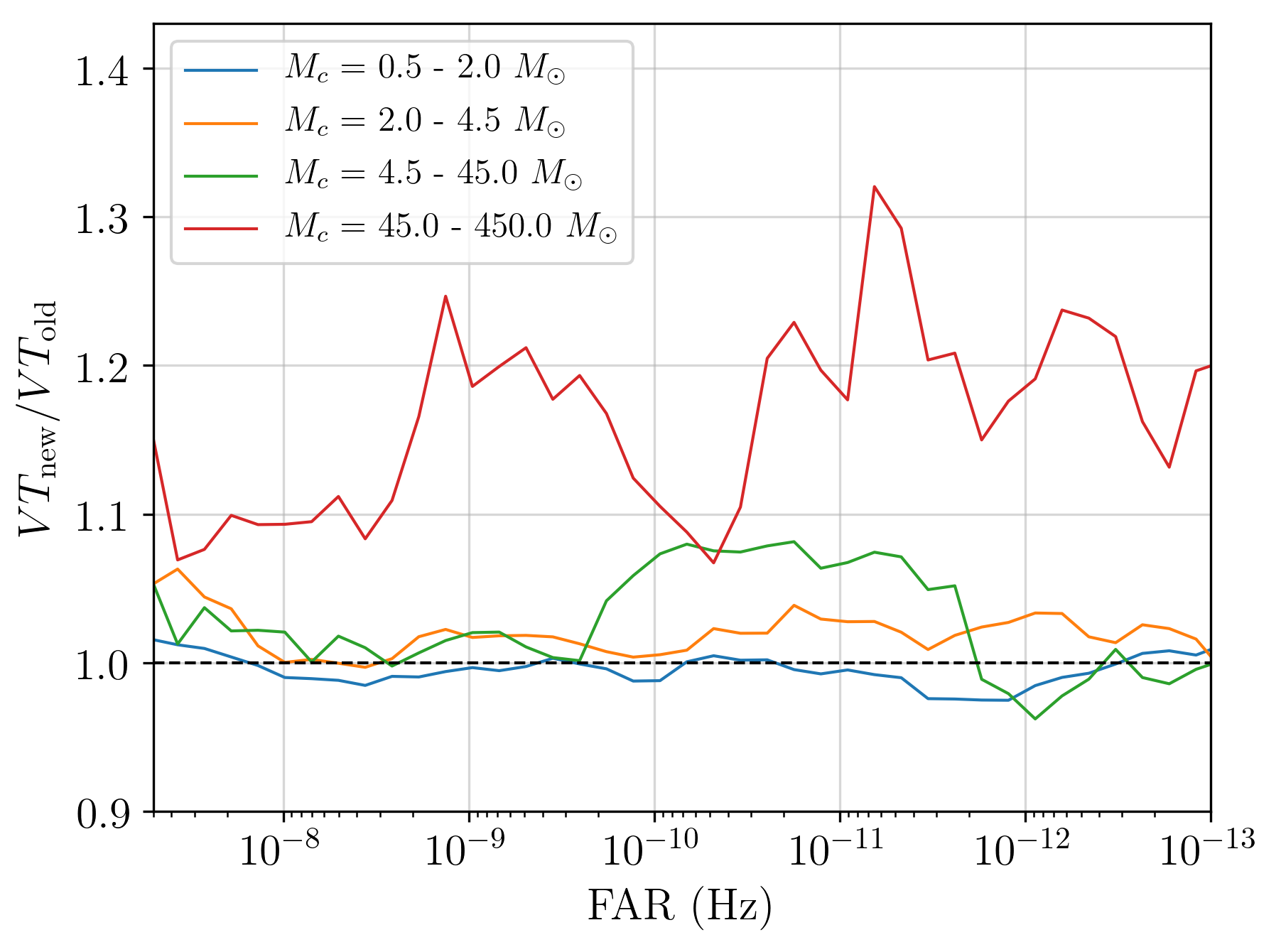}
    \caption{\label{fig:combinedchisq_vtratio}
    Ratio of the \ac{vt} value measured by the run with the combined $\xi^2$ to
    that with the normal $\xi^2_\mathrm{a}$ as a function of \ac{far}, and hence
    the value above 1 indicates the sensitivity improvement due to the
    incorporation of bank $\xi^2$ statistics. The different colors represent the
    four chirp-mass ($M_c$) bins mentioned in the legend.
    }
\end{figure}

Given the injection set, we conduct two sets of the simulated observations with
the conventional $\xi_\mathrm{a}^2$ and the combined $\xi^2_\mathrm{ab}$
statistics defined in \eqref{eq:combinedchisq_def}, respectively, and measure
the \ac{vt} value for each of the two cases.  When evaluating the \ac{vt}
values, we bin the entire set of injections into four chirp-mass ($M_c$) ranges:
\SIrange{0.5}{2.0}{}$\ M_\odot$(\ac{bns}), \SIrange{2.0}{4.5}{}$\
M_\odot$(lighter \ac{bbh}), \SIrange{4.5}{45}{}$\ M_\odot$(heavier \ac{bbh}) and
\SIrange{45}{450}{}$\ M_\odot$(\ac{imbh}). \figref{fig:combinedchisq_vtratio} is
a ratio of the \ac{vt} value measured by the run with the combined
$\xi^2_\mathrm{ab}$ to that with the $\xi^2_\mathrm{a}$ as a function of
\ac{far}, and hence the values above 1 indicates the sensitivity improvement due
to the incorporation of bank-$\xi^2$ statistics. The different colors represent
the four chirp-mass bins mentioned above.

One can find that at the \ac{far} of \SI{3.2e-8}{\hertz}($\approx1$ per year)
the \ac{vt} ratio increases by \SI{10}{\percent} (or even more at lower
\ac{far}) for \ac{imbh} injections, while the other three categories do not
exhibit noticeable improvement. This difference can be understood by the
duration of those injected signals in the detector's frequency band. Given the
shortest duration of \ac{imbh} signals, the time-domain consistency test
performed by the $\xi^2_\mathrm{a}$ statistic do not help those signals to be
distinguished from noise, and hence the complimentary test on the template
domain using the $\xi^2_\mathrm{ab}$ statistic is rather informative, leading to
the significant improvement in the \ac{vt} value as shown in
\figref{fig:combinedchisq_vtratio}. Note that the zig-zaggy structure in
\ac{imbh}'s \ac{vt} ratio is due to relatively large uncertainty in each \ac{vt}
measurement given a smaller number of recovered \ac{imbh} injections than other
source categories.
\subsection{\label{sec:results_newsignalmodel} Upgraded $\rho-\xi^2$ signal model}
\begin{figure}[htbp]
    \includegraphics[width=\linewidth]{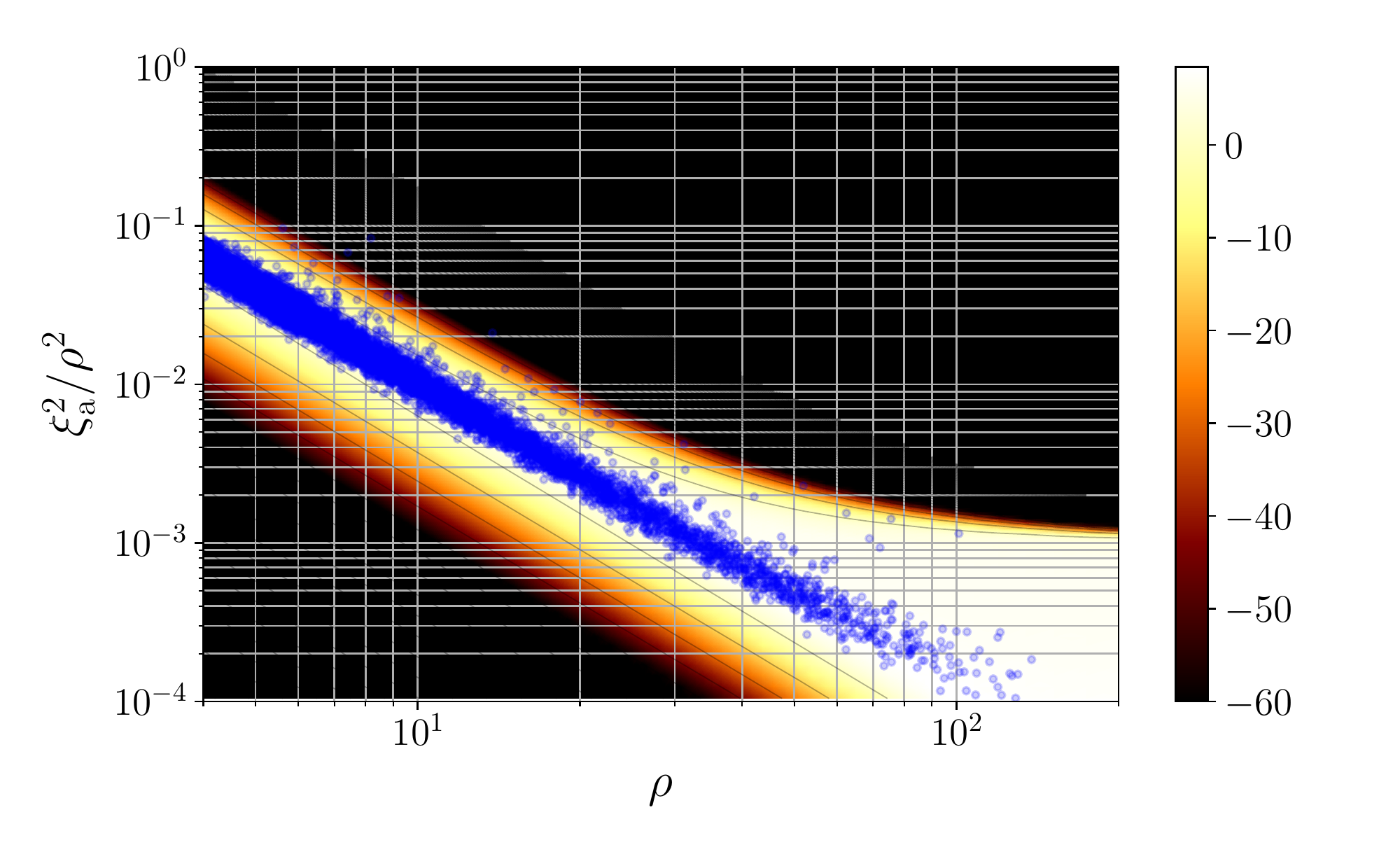}
    \caption{\label{fig:sig_scatter}
    Scatter plot of recovered \ac{bns} injections, denoted by blue circles, on
    top of the $P(\xi^2_H \mid \rho_H, \{\bar{\theta}\},
    \mathcal{H}_\mathrm{s})$ associated with \ac{bns} templates for \ac{ligo}
    Hanford detector.  The color scheme encodes the probability density in a
    logarithmic scale with a brighter region having larger values.
    }
\end{figure}

We analyze the same dataset and injection set described in
\secref{sec:results_combinedchisq}. Yet, we use the template bank developed for
\ac{o4}, adopting the \texttt{manifold} placement algorithm~\cite{manifold} and
the sorting scheme using the orthogonalized PN-phase terms described
in~\cite{shio_template_bank, morisaki_froq}.  \figref{fig:sig_scatter} shows a scatter plot of
recovered \ac{bns} injections, represented by blue circles, on top of the $P(\xi^2_H
\mid \rho_H, \{\bar{\theta}\}, \mathcal{H}_\mathrm{s})$ \ac{pdf} associated with one of
the \ac{bns} template groups for \ac{ligo} Hanford detector. This visually
demonstrates that the density of recovered \ac{bns} injections on $(\rho,
\xi^2)$ parameter space is largely consistent with the upgraded signal model,
verifying the validity of the derivation described in
\secref{sec:chisq_signal_model}.

\begin{figure}[htbp]
    \includegraphics[width=\linewidth]{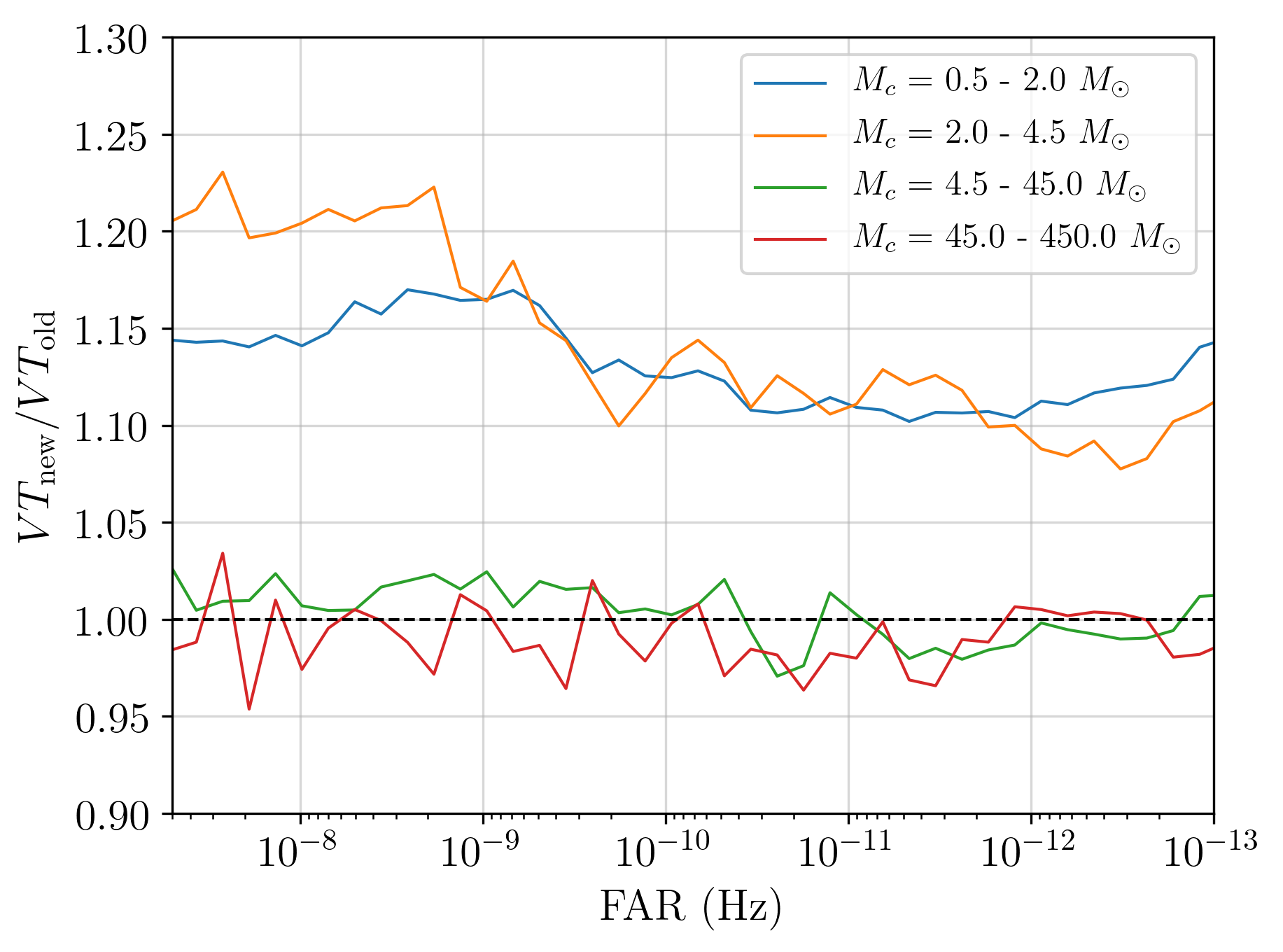}
    \caption{\label{fig:sig_vt_ratio}
    Ratio of the \ac{vt} value measured by the run with the upgraded signal
    model to that with the original signal model as a function of \ac{far}, and
    hence the values above 1 indicates the sensitivity improvement due to the
    upgraded signal model. The different colors represent the four chirp-mass ($M_c$)
    bins mentioned in the legend. 
    }
\end{figure}

Given the injection set, we conduct two sets of the simulated observations with
and without the upgraded $\rho-\xi^2$ signal model, respectively, and measure
the \ac{vt} value for each of the two cases. Binning the injection set in the
same way as \secref{sec:results_combinedchisq}, \figref{fig:sig_vt_ratio} shows a
ratio of the \ac{vt} value measured by the run with the upgraded signal model to
that with the original signal model as a function of \ac{far}. This figure implies that
at the \ac{far} of \SI{3.2e-8}{\hertz}($\approx1$ per year) the \ac{vt} ratio
increases by \SI{15}{\percent} (\SI{20}{\percent}) for \ac{bns} (lighter
\ac{bbh}) injections, while the heavier \ac{bbh} and \ac{imbh} injections do not
exhibit noticeable improvement.  Since
\ac{bns} and lighter \ac{bbh} signals produce longer duration, the time-domain
consistency test by $\xi^2_\mathrm{a}$ statistics tend to be more
impactful for these source categories than heavier ones.
\section{Conclusion}
\label{sec:concl}
In this work we have described several new features implemented in
\gstlal{}-based insprial pipeline, leading up to \ac{o4}.  These features
consist of: the signal contamination removal, the bank-$\xi^2$ incorporation,
the upgraded $\rho-\xi^2$ signal model and the integration of KAGRA.
Specifically, we have demonstrated by the \ac{vt} comparison that the
bank-$\xi^2$ incorporation improves the sensitivity to \ac{imbh} signals by
\SI{10}{\percent} or more and that the upgraded $\rho-\xi^2$ signal model
improves the sensitivity to \ac{bns} (lighter \ac{bbh}) signals by
\SI{15}{\percent} (\SI{20}{\percent}), respectively.  Although we have not
quantitatively shown the performance of the signal contamination removal,
\figref{fig:signal_contamination} visually illustrates that the signal
contamination can be fully removed. A more thorough investigation using blind
injections is described in~\cite{count_tracker}. Also, in the injection study
shown in \secref{sec:results}, we did not incorporate the upgraded
$\rho-\xi^2$ signal model and the bank-$\xi^2$ statistic both simultaneously for one
configuration, which requires derivation and recalculation of the covariance
matrix shown in \eqref{eq:autochisq_covmat}. We leave this as future work to
address during offline reanalysis of \ac{o4} dataset. Regarding the overall
performance of the latest \gstlal{} analysis, see~\cite{o4_performance} for the
detailed results of the Mock Data Challenge conducted as preparation of \ac{o4}.

\begin{acknowledgments}
    The authors are grateful for computational resources provided by the LIGO
    Laboratory and supported by National Science Foundation Grants PHY-0757058
    and PHY-0823459.  This material is based upon work supported by NSF's LIGO
    Laboratory which is a major facility fully funded by the National Science
    Foundation.  LIGO was constructed by the California Institute of Technology
    and Massachusetts Institute of Technology with funding from the National
    Science Foundation (NSF) and operates under cooperative agreement
    PHY-1764464.  The authors are grateful for computational resources provided
    by the Pennsylvania State University's Institute for Computational and Data
    Sciences (ICDS) and the University of Wisconsin Milwaukee Nemo and support
    by NSF PHY-\(2011865\), NSF OAC-\(2103662\), NSF PHY-\(1626190\), NSF
    PHY-\(1700765\), NSF PHY-\(2207728\), and NSF PHY-\(2207594\).  This paper
    carries LIGO Document Number LIGO-P2300116.

    This research has made use of data or software obtained from the
    Gravitational Wave Open Science Center (gwosc.org), a service of LIGO
    Laboratory, the LIGO Scientific Collaboration, the Virgo Collaboration, and
    KAGRA. LIGO Laboratory and Advanced LIGO are funded by the United States
    National Science Foundation (NSF) as well as the Science and Technology
    Facilities Council (STFC) of the United Kingdom, the Max-Planck-Society
    (MPS), and the State of Niedersachsen/Germany for support of the
    construction of Advanced LIGO and construction and operation of the GEO600
    detector. Additional support for Advanced LIGO was provided by the
    Australian Research Council. Virgo is funded, through the European
    Gravitational Observatory (EGO), by the French Centre National de Recherche
    Scientifique (CNRS), the Italian Istituto Nazionale di Fisica Nucleare
    (INFN) and the Dutch Nikhef, with contributions by institutions from
    Belgium, Germany, Greece, Hungary, Ireland, Japan, Monaco, Poland, Portugal,
    Spain. KAGRA is supported by Ministry of Education, Culture, Sports, Science
    and Technology (MEXT), Japan Society for the Promotion of Science (JSPS) in
    Japan; National Research Foundation (NRF) and Ministry of Science and ICT
    (MSIT) in Korea; Academia Sinica (AS) and National Science and Technology
    Council (NSTC) in Taiwan.
\end{acknowledgments}

\appendix
\section{Derivation of \eqref{eq:jacobian_det}}
\label{app:jacobian}
We start with the following coordinate transformation shown in
\eqref{eq:dtdphi_coodtrans}:
\begin{align}
\label{eq:cood_trans}
\vec{\rho} \rightarrow
\vec{x}=
\left(
\begin{array}{c}
\rho_\mathrm{net} \\
\ln\left\{\mathcal{D}_{2}/\mathcal{D}_\mathrm{ref}\right\} \\
\vdots \\
\ln\left\{\mathcal{D}_{n}/\mathcal{D}_\mathrm{ref}\right\}
\end{array}
\right).
\end{align}
In general, any coordinate transformation in the probability density involves
modification in the volume element characterized by Jacobian matrix, which explicitly reads
\begin{align}
    \label{eq:jacobian_mat}
\boldsymbol{\mathcal{J}}(\vec{\rho})=
\begin{pmatrix}
\frac{\partial x_1}{\partial \rho_1} & \cdots & \frac{\partial x_i}{\partial \rho_1} & \cdots & \frac{\partial x_n}{\partial \rho_1}\\
\vdots & \ddots &        &        & \vdots \\
\frac{\partial x_1}{\partial \rho_j} &        & \frac{\partial x_i}{\partial \rho_j} &        & \frac{\partial x_n}{\partial \rho_j} \\
\vdots &        &        & \ddots & \vdots \\
\frac{\partial x_1}{\partial \rho_n} & \cdots & \frac{\partial x_i}{\partial \rho_n} & \cdots & \frac{\partial x_n}{\partial \rho_n}
\end{pmatrix}
.
\end{align}
Note that in this particular case the new coordinates, $\vec{x}$,
can be written in terms of the original \ac{snr} coordinates such that
\begin{align}
    x_1 = \rho_\mathrm{net},\qquad x_i = \ln\left\{\frac{D_i}{D_1}\right\} - \ln\left\{\frac{\rho_i}{\rho_1}\right\},
\end{align}
where $D_i$ is the horizon distance for $i$-th detector.

For example, when evaluating \eqref{eq:dtdphi} for coincident triggers
between \ac{ligo} Hanford and Livingston detectors, the SNR coordinates are
given by
\begin{align}
    \rho_1=\rho_H,\qquad \rho_2 = \rho_L.
\end{align}
Hence, the Jacobian matrix \eqref{eq:jacobian_mat} takes the form of 
\begin{align}
\boldsymbol{\mathcal{J}}(\rho_H, \rho_L)=
\begin{pmatrix}
\frac{\rho_H}{\sqrt{\rho_H^2+\rho_L^2}} & &  \frac{\rho_L}{\sqrt{\rho_H^2+\rho_L^2}}\\
 & & \\
\frac{1}{\rho_H} & & -\frac{1}{\rho_L} \\
\end{pmatrix}
,
\end{align}
which leads to the determinant
\begin{align}
    |\boldsymbol{\mathcal{J}}(\vec{\rho})|=\frac{\sqrt{\rho_H^2 + \rho_L^2}}{\rho_H\rho_L}.
\end{align}
Similarly, this derivation can be extended to the three-detector case, and in
general, one finds the general expression of the determinant shown in
\eqref{eq:jacobian_det}.
\section{Derivation of \eqref{eq:p_of_t_noise}}
\label{app:trig_rate}
The probability of observing $k$ noise events with the mean rate $\mu$ during unit
observation time $T$ follows a Poisson distribution
\begin{align}
    \label{eq:poisson_dist}
    P(N=k\mid \mu) = \frac{1}{k!}(\mu T)^k \eexp^{-\mu T}.
\end{align}
In \gstlal's implementation, the \textit{total} event rate consists of several categories
characterized by a template group $\{\bar{\theta}\}$, i.e. $\mu =
\sum_{\{\bar{\theta}\}}\mu_{\{\bar{\theta}\}}$, and \eqref{eq:poisson_dist} is
assumed to hold independently for noise events from each category. Since
$P\left(t_\mathrm{ref}, \theta \mid \mathcal{H}_\mathrm{n}\right)$ considers a
situation where the given event occurs for the associated template group and no
others, the \ac{pdf} reads
\begin{align}
    P\left(t_\mathrm{ref}, \theta \mid \mathcal{H}_\mathrm{n}\right) &= P(1\mid \mu_{\{\bar{\theta}\}})\prod_{i\neq \{\bar{\theta}\}} P(0\mid \mu_i)\\
    &= \mu_{\{\bar{\theta}\}} T \eexp^{-\mu T}.
\end{align}
Since the exponential factor is constant across different template groups, the
only dependence on $\theta$ boils down to $\mu_{\{\bar{\theta}\}}$ as shown in
\eqref{eq:p_of_t_noise}.

\section{Derivation of \eqref{eq:Usq}}
\label{app:chisq_derive}
From the definition of $\boldsymbol{\Sigma}$ shown in \eqref{eq:sigma_def}, one
can find that $\boldsymbol{\Sigma}$ is positive semidefinite, and hence, there
exists square root of $\boldsymbol{\Sigma}$ such that
$\boldsymbol{A}^2=\boldsymbol{\Sigma}$.  Introducing
$\vec{V}=\boldsymbol{A}^{-1}\vec{U}$, from $\vec{U}\sim\mathcal{N}(\vec{\mu},
\boldsymbol{\Sigma})$, it follows that
$\vec{V}\sim\mathcal{N}(\boldsymbol{A}^{-1}\vec{\mu}, \mathbbm{1})$.
Furthermore, since $\boldsymbol{\Sigma}$ can be diagonalized as
$\boldsymbol{\Sigma}=\boldsymbol{P}^\dag\boldsymbol{\Lambda}\boldsymbol{P}$,
where $\boldsymbol{P}$ is a unitary transformation matrix and
$\boldsymbol{\Lambda}$ is a diagonal matrix with entries $\lambda_i$,
\eqref{eq:Usq} reads
\begin{align}
    \vec{U}^\dag\vec{U} &= \vec{V}^\dag(\boldsymbol{A}^2)\vec{V} = \vec{V}^\dag(\boldsymbol{P}^\dag\boldsymbol{\Lambda}\boldsymbol{P})\vec{V}\\
    \label{eq:Usq_3}
    &= \vec{W}^\dag\boldsymbol{\Lambda}\vec{W} = \sum_i\lambda_i|\hat{n}_i+\eta_i|^2.
\end{align}
Here \eqref{eq:Usq_3} is given by the fact that
$\vec{W}=\boldsymbol{P}\vec{V}\sim\mathcal{N}(\vec{\eta}, \mathbbm{1})$, where
$\vec{\eta}=\boldsymbol{A}^{-1}\vec{\mu}$. This also implies that $|\hat{n}_i+\eta_i|^2$
obeys noncentral chi-square distribution with the 2 degrees of freedom and
noncentral parameter of $|\eta_i|^2$. Therefore, this formalism suggests that
$\vec{U}^\dag\vec{U}$, or equivalently the $\xi^2$ parameter, is a weighted sum
(characterized by $\lambda_i$) of multiple chi-square random variables, which is
in general referred to as \textit{generalized chi-square distribution}.
See~\cite{mathai1992quadratic} for more detailed discussion.
\normalem
\bibliography{references}% Produces the bibliography via BibTeX.

\end{document}